\def \AAP #1 #2 {{\em A{\&}A\/} {\bf #1}, #2}  
\def \AAL #1 #2 {{\em AA{\&}A Lett.\/} {\bf #1}, L#2}  
\def \AAR #1 #2 {{\em A{\&}A Rev.\/} {\bf #1}, #2}  
\def \AAS #1 #2 {{\em A{\&}A Suppl. Ser.\/} {\bf #1}, #2}  
\def \AJ #1 #2 {{\em AJ\/} {\bf #1}, #2}  
\def \ANNREV #1 #2 {{\em Ann. Rev. Astron. Astrophys.\/} {\bf #1}, #2}  
\def \APJ #1 #2 {{\em ApJ\/} {\bf #1}, #2}  
\def \APJL #1 #2 {{\em ApJ Lett.\/} {\bf #1}, L#2}  
\def \APJS #1 #2 {{\em ApJ Suppl.\/} {\bf #1}, #2}  
\def \APSS #1 #2 {{\em Astrophys. Space Sci.\/} {\bf #1}, #2}  
\def \ASR #1 #2 {{\em Adv. Space Res.\/} {\bf #1}, #2}  
\def \MN #1 #2 {{\em MNRAS\/} {\bf #1}, #2}  
\def \PRL #1 #2 {{\em Phys. Rev. Lett.\/} {\bf #1}, #2}  
\def \PRD #1 #2 {{\em Phys. Rev. D\/} {\bf #1}, #2}
\def \NAT #1 #2 {{\em Nature\/} {\bf #1}, #2}   
\def \SCI #1 #2 {{\em Science\/} {\bf #1}, #2}   
  
\documentclass{elsart3p}
\usepackage{natbib}
\usepackage{amssymb}
\usepackage{epsfig}
 
\begin{document}

\begin{frontmatter}
  
\title{Cosmic ray acceleration at relativistic shocks, shear layers, ...}  
  
\author{Micha{\l} Ostrowski}  
  
\address{Obserwatorium Astronomiczne, Uniwersytet Jagiello\'nski \\ 
ul. Orla 171, 30-244 Krak\'ow, Poland \\(E-mail:  mio{@}oa.uj.edu.pl)}

\begin{abstract}  
A short discussion of theoretical results on cosmic ray first-order Fermi acceleration at relativistic shock waves is presented. We point out that the recent results by Niemiec with collaborators change the knowledge about these processes in a substantial way. In particular one can not expect such shocks to form particle distributions extending to very high energies. Instead, distributions with the shock compressed injected component followed by a more or less extended high energy tail are usually created. Increasing the shock Lorentz factor leads to steepening and decreasing of the energetic tail. Also, even if a given section of the spectrum preserves the power-law form, the fitted spectral index may be larger or smaller than the claimed `universal index' $\sigma \approx 2.2$~. 

A reported simple check of real shapes of electron spectra in the Cyg A hot spots provides results clearly deviating from the standard expectations for such shocks met in the literature. The spectrum consist of the very flat low energy part ($\sigma \approx 1.5$), up to electron energies $\sim 1$ GeV, and much steeper part ($\sigma > 3$) at higher energies. We conclude with presentation of a short qualitative discussion of the Fermi second-order processes acting in relativistic plasmas. We suggest that such processes can be the main accelerating agent for very high energy particles. In particular its can accelerate electrons to energies in the range of $1$ - $10^3$ TeV in  relativistic jets, shocks and radio-source lobes.
\end{abstract} 


\end{frontmatter}
 
\section{Introduction}  
  
Relativistic plasma flows are observed in a number of astrophysical objects, ranging from a mildly relativistic jets of  the sources like SS433, through the-Lorentz-factor-of-a-few jets in AGNs and galactic  `micro-quasars', up to highly ultra-relativistic outflows in sources of gamma  ray bursts or pulsar winds. As nearly all such objects are efficient emitters of nonthermal radiation, what requires existence of energetic particles, our attempts to  understand the processes generating cosmic rays are essential for understanding the fascinating phenomena observed. Below I will discuss the work carried out in order to understand the cosmic ray acceleration processes acting at relativistic shocks and within highly turbulent regions accompanying such shocks and shear layers. I will not include here the interesting work involving collisionless shocks modelling with {\it particle in cell} simulations. This approach uses quite different modelling method in comparison to the other work discussed here,  relating in most cases to the characteristic energy range of shock compressed thermal plasma particles.  

The present paper is a modified and updated version of some of my previous reviews of the subject. Also, essentially the same slightly shortened text is provided as my contribution to the HEPRO Workshop in Dublin (September 2007).

Below we will append the index `1' (`2') to quantities measured in the plasma rest frame upstream (downstream) of the shock.
  
\section{Particle diffusive acceleration at non-relativistic shock waves}  
  
Processes of the first-order particle acceleration at non-relativistic shock waves were widely discussed by a number of authors, let me note still actual reviews by Drury (1983) and Blandford \& Eichler (1987). The most interesting physical feature of the first-order shock acceleration at the non-relativistic shock wave is independence of the {\it test-particle stationary} particle energy spectrum from the background  conditions near the shock, including the mean magnetic field  configuration and the spectrum of MHD turbulence. The reason is a nearly-isotropic form of the particle momentum distribution at the shock. If efficient scattering occurs near the shock, this condition also holds for oblique shocks with the shock velocity along the (upstream) magnetic field $U_{B,1} \equiv U_1 / \cos \Psi_1 \ll v$ ($\Psi_1$ - the upstream magnetic field inclination to the shock normal). Then, the particle density is  continuous across the shock and the spectral index for the {\em phase-space} distribution function, $\alpha$, is given exclusively in terms of a single parameter -- the shock compression ratio $R$:  
  
$$\alpha = {3R \over R-1}  \qquad   . \eqno(2.1) $$  
  
\vspace{1mm}
\noindent  
Because of the isotropic form of the particle distribution function, the  
spatial diffusion equation has become a widely used mathematical tool  
for describing particle transport and acceleration processes in  
non-relativistic flows. 
  
In real astronomical objects some non-linear and time dependent effects, or occurring of additional energy losses and gains can make the physics of the  
acceleration more complicated, creating, e.g., non-power-low and/or  
non-stationary particle distributions.

\section{Cosmic ray acceleration at relativistic shock waves}  

Below I describe work done on mildly- and ultra-relativistic shock acceleration including important recent results of Niemiec et al. With these last results many previous ones occurred to be of historical value only, reflecting specific individual features of the acceleration processes. Basing on these older works one can understand the recent simulations results in a relatively straightforward way. Attempting to give an overview of the full field I base this presentation on my own and my collaborators work, which seems to me to present a consistent way of development and reflects my approach to understanding acceleration processes in relativistic shocks.

\subsection{History: acceleration at mildly relativistic shocks}  
  
In cases of the shock velocity reaching values comparable to the light velocity, the particle distribution at the shock becomes {\em anisotropic}. This simple fact complicates to a great extent both the physical picture and the mathematical description of particle acceleration. A first attempt to consider the acceleration process at the relativistic shock was presented in 1981 by Peacock, and a consistent theory was proposed later by Kirk \& Schneider (1987a). Those authors considered stationary solutions of the relativistic Fokker-Planck equation for particle pitch-angle diffusion in the parallel shock wave. In the situation with the gyro-phase averaged distribution $f(p, \mu, z)$, which depends only on the unique spatial co-ordinate $z$ along the shock velocity, and with $\mu$ being the pitch-angle cosine, the equation takes a form:  
  
$$\Gamma ( U + v \mu ) {\partial f \over \partial z} = C(f) + S \qquad ,  
 \eqno(3.1)  $$  
  
\vspace{1mm}
\noindent  
where $\Gamma \equiv 1/\sqrt{1-U^2}$ is the flow Lorentz factor, $C(f)$  
is the collision operator and $S$ is the source function. In the  
presented approach, the spatial co-ordinates are measured in the shock  
rest frame, while the particle momentum co-ordinates and the collision  
operator are given in the respective plasma rest frame. For the applied  
pitch-angle diffusion operator, $C = \partial / \partial \mu (D_{\mu  
\mu} \partial f / \partial \mu)$, the authors generalized the diffusive  
approach to higher order terms in particle distribution anisotropy and  
constructed general solutions at both sides of the shock which involved  
solutions of the eigenvalue problem. By matching two solutions at the  
shock, the spectral index of the resulting power-law particle  
distribution can be found by taking into account a sufficiently large  
number of eigenfunctions. The same procedure yields the particle angular  
distribution and the spatial density distribution. The low-order  
truncation in this approach corresponds to the standard diffusion  
approximation and to a somewhat more general method described by  
Peacock.  
   
An application of this approach to more realistic conditions -- but  
still for parallel shocks -- was presented by Heavens \& Drury (1988),  
who investigated the fluid dynamics of relativistic shocks (cf. also  
Ellison \& Reynolds 1991) and used the results to calculate spectral  
indices for accelerated particles. 
  
\begin{figure}[hbt]  
\vspace{56mm}
\includegraphics{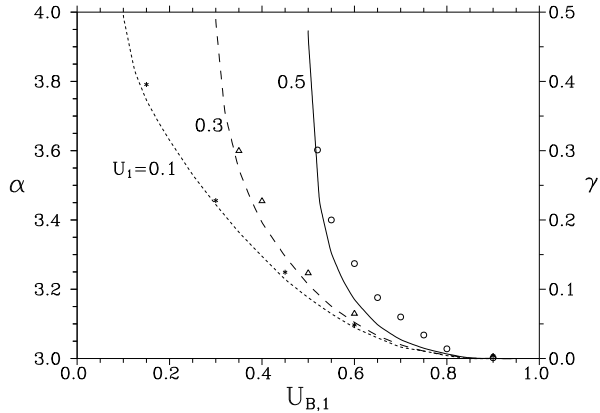}   
\vspace{0mm}
\caption{ 
Spectral indices $\alpha$ of particles accelerated at oblique shocks  
versus shock velocity projected at the mean magnetic field, $U_{B,1}$ (Ostrowski 1991a). The results are presented for the shock compression $R = 4$. On the right the respective synchrotron spectral index $\gamma$ is given. The shock velocities $U_1$ are given near the respective curves taken from Kirk \& Heavens (1989). The points were taken from simulations computing explicitly the details of particle-shock interactions (Ostrowski 1991a). 
} 
\end{figure}  
  
A substantial progress in understanding the acceleration process in the  
presence of highly anisotropic particle distributions is due to the work  
of Kirk \& Heavens (1989; see also Ostrowski 1991a and Ballard \&  
Heavens 1991), who considered particle acceleration at {\it subluminal}  
($U_{B,1} < c$) relativistic shocks with oblique magnetic fields. They  
assumed the magnetic momentum conservation, $p_\perp^2/B = const$, at  
particle interaction with the shock and applied the Fokker-Planck  
equation discussed above to describe particle transport along the field  
lines outside the shock, while excluding  the possibility of cross-field  
diffusion. In the cases when $U_{B,1}$ reached relativistic values, they  
derived very flat energy spectra with $\gamma \approx 0$ at $U_{B,1}  
\approx 1$ (Fig.~1). In such conditions, the particle density in front of the shock can substantially -- even by a few orders of magnitude -- exceed  
the downstream density (see the curve denoted `-8.9' at Fig.~2).  
Creating flat spectra and great density contrasts is possible due to the effective repeating reflections of anisotropically distributed upstream particles from the region of compressed magnetic field downstream of the shock. 
  
As stressed by Begelman \& Kirk (1990), in relativistic shocks one  
often finds {\it superluminal} conditions with $U_{B,1} > c$, where  
the above presented approach is no longer valid. Then, it is not  
possible to reflect upstream particles from the shock and to transmit  
downstream particles into the upstream region. In effect, only a single  
transmission of upstream (or shock injected) particles re-shapes the original distribution by shifting particle energies to larger values, with super-adiabatic efficiency due to anisotropic particle distribution at the transmission.   

\begin{figure}[hbt]  
\vspace{74mm}
\includegraphics{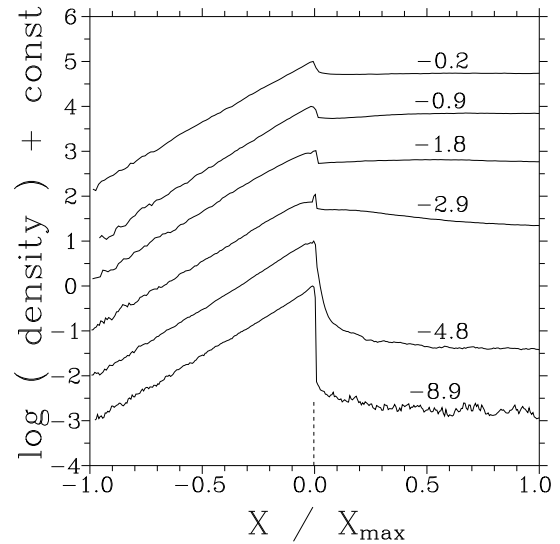}   
\vspace{0mm}
\caption{ 
Energetic particle density profiles across the relativistic shock  
with the oblique magnetic field (Ostrowski 1991b). The shock with $U_1 =  
0.5$, $R = 5.11$ and $\psi_1 = 55^o$ is considered. The curves for  
growing to the top perturbation amplitudes are characterized with the value $\log \kappa_\perp / \kappa_\parallel$ ($\kappa_\perp / \kappa_\parallel$ is a ratio of the cross-field to the parallel diffusion coefficients) given near each curve. The data are vertically shifted for picture clarity. The value $X_{max}$ is the distance from the shock at which the upstream particle density decreases  
to $10^{-3}$ part of the shock value.} 
\end{figure}

\subsection{History: acceleration in the presence of large amplitude  
magnetic field perturbations}

As the approaches proposed by Kirk \& Schneider (1987a) and Kirk \& Heavens  
(1989), and the derivations of Begelman \& Kirk (1990) are valid only in  
cases of weakly perturbed magnetic fields, for larger amplitude MHD perturbations numerical methods have to be used. A series of such simulations were performed by numerous authors (e.g. Kirk \& Schneider 1987b; Ellison et al. 1990; Ostrowski 1991a, 1993; Ballard \& Heavens 1992, Naito \& Takahara 1995, Bednarz \&  Ostrowski 1996). Different approaches applied, including the ones involving particle momentum pitch-angle diffusion or integrating particle trajectories in analytic structures of the perturbed magnetic fields\footnote{Let us note that application by some authors of point-like large-angle scattering models in relativistic shocks does not provide a viable physical representation of the scattering at MHD waves (Bednarz \& Ostrowski 1996).}, allowed for considering a wide range of background conditions near the shock. The results obtained by different authors can be summarized at the figure (Fig.~3) taken from Bednarz \& Ostrowski (1996). One should note, that essentially all derivations by the above authors were performed with assuming scale-free conditions for the acceleration process, resulting in power-law distributions of the accelerated particles. 
 
\begin{figure}[hbt]  
\vspace{57mm}
\includegraphics{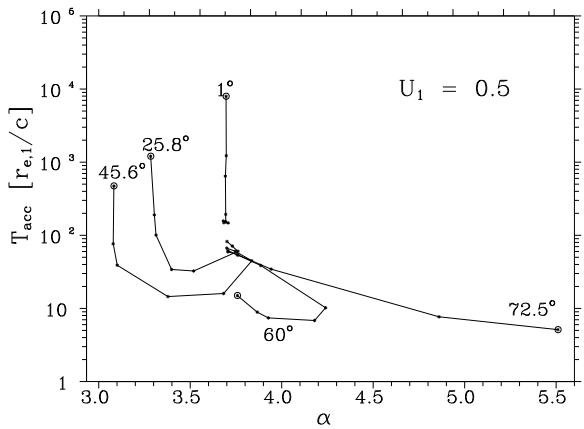}   
\vspace{0mm}
\caption{ 
Relation of the particle acceleration time scale $T_{acc}$ versus the particle spectral index $\alpha$ at different magnetic field inclinations $\psi_1$ given near the respective curves (Bednarz \& Ostrowski 1996). The {\it minimum} value of the model parameter $\kappa_\perp/\kappa_\|$ occurs at the encircled point of each curve and the wave amplitude monotonously increases along each curve up to $\delta B \sim B$, where all curves converge in $\alpha$. The curve for $\Psi_1 = 60^\circ$ ($U_{B,1}=1$) separate the sub- and super-luminal shock results. We do not discuss here the presented acceleration times.
} 
\end{figure}  
  
At the figure (Fig. ~3) one can find very flat spectra for oblique subluminal shocks if the perturbation amplitudes are small. Contrary to that generation of the power-law spectrum is possible in the superluminal shocks only in the presence of large amplitude turbulence. Then, in contrast to the subluminal shocks, spectra are extremely steep for small values of $\delta B$ (not presented at the figure) and $\alpha$ monotonously decreases with increasing  magnetic field perturbations. A new feature is observed in oblique shocks of the spectral index $\alpha$ changing with $\delta B$ in a non-monotonic way. 
  
\subsection{History: Energy spectra of cosmic rays accelerated at ultra-relativistic shocks}  
  
\noindent  
The main difficulty in modelling the acceleration processes at shocks with  
large Lorentz factors $\Gamma$ is the fact that the involved particle  
distributions are extremely anisotropic in the shock, with the  
particle angular distribution opening angles $\sim \Gamma^{-1}$ in the  
upstream plasma rest frame. In the simulations of Bednarz \& Ostrowski (1998) a Monte Carlo method involving small amplitude pitch-angle scattering was applied for particle transport near the shocks with $\Gamma$ in the range $3$ -- $243$. The simulations revealed an unexpected result, showing convergence, for $\Gamma \to \infty$, of the resulting power-law distributions to the 'universal' one with the spectral index $\sigma \approx 2.2$ (Fig.~4). Essentially the same result was derived with different methods by many other authors (Gallant \& Achterberg 1999; Kirk et al. 2000, Achterberg et al. 2001, Lemoione et al. 2003, Keshet \& Waxman 2005, Lemoine \& Revenu 2006, Morlino et al. 2007, et al.), what could suggest that in the ultrarelativistic limit the acceleration process becomes again simple, generating cosmic ray spectra essentially independent of the considered background conditions. One can find examples in the literature that some authors extend this claim for all relativistic shocks.

\begin{figure}[hbt]  
\vspace{57mm}
\includegraphics{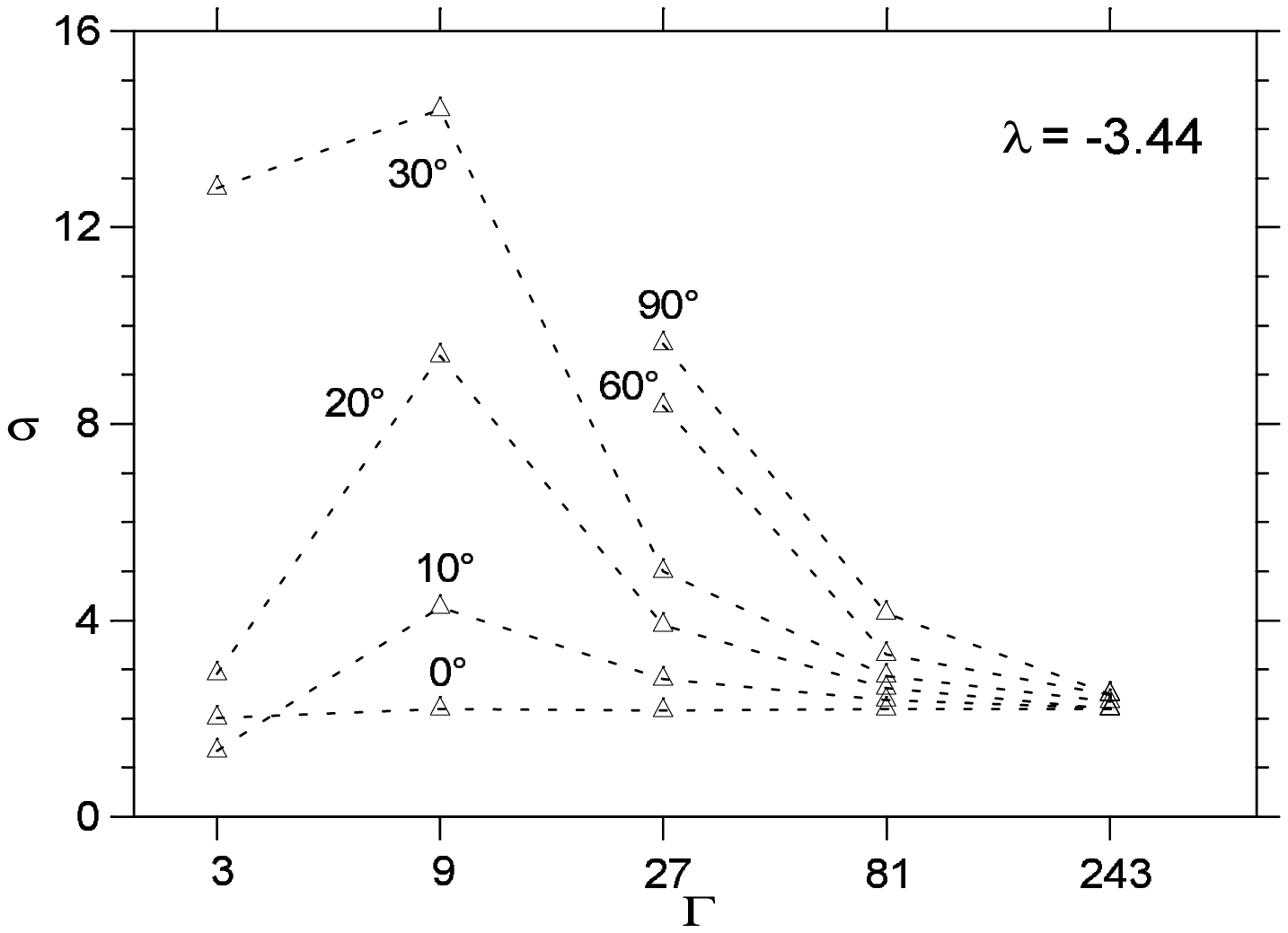}  
\vspace{0mm}
\caption{ 
The simulated spectral indices $\sigma$ ($\sigma \equiv \alpha - 2$) versus the shock Lorentz factor $\Gamma$ (Bednarz \& Ostrowski 1998). Results for a given $\psi_1$ are joined with dashed lines; the respective value of $\psi_1$ is given near each curve.  Increasing the turbulence amplitude in a not presented here series of simulations led to shifting the resulting curves toward the parallel shock, $\Psi_1 = 0^\circ$, results.} 
\end{figure}     
  
\begin{figure*}[hbt]  
\vspace{73mm}
\includegraphics{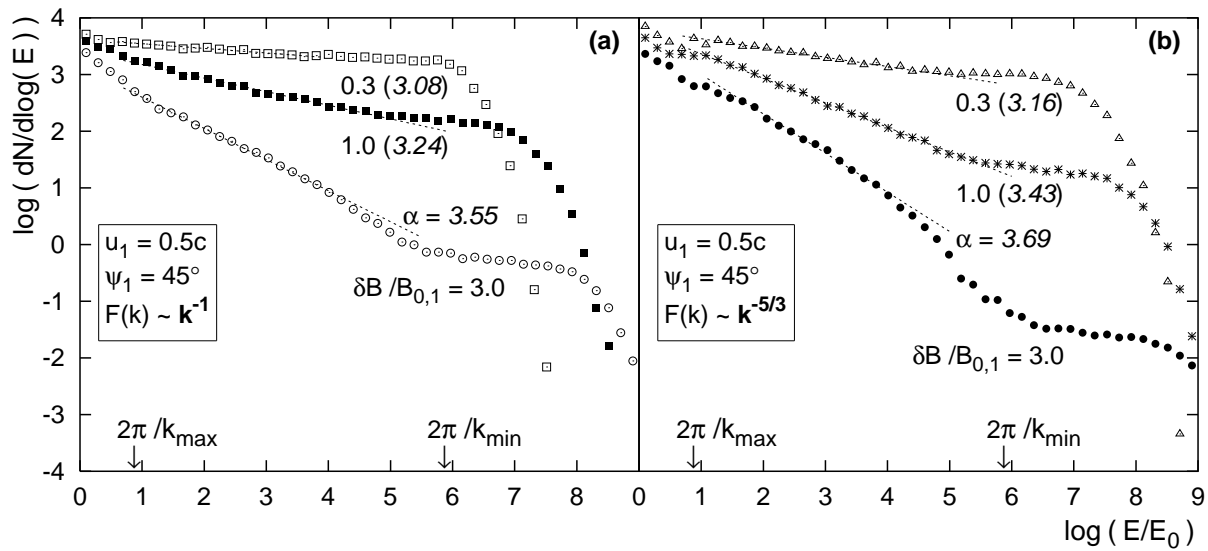}   
\vspace{0mm}
\caption{ 
Particle spectra for an oblique mildly relativistic shock: the shock velocity $u_1 = 0.5\,c$, the mean magnetic field inclination $\Psi_1 = 45^\circ$ and the wave power spectrum are indicated in the respective panels (from Niemiec \& Ostrowski 2004). Values of magnetic turbulence amplitude, $\delta B/B$, and the indices fitted to the power-law sections of spectra (in parentheses) are given near each result.} 
\end{figure*}

Ostrowski \& Bednarz (2002) reconsidered all the above approaches to derive  particle spectra at relativistic shocks and `discovered' that the conditions producing the universal spectral index were in some way equivalent to assuming highly turbulent conditions near the shock. Additionally, all these models did not introduce any physical scale and thus forced the power-law shape of the resulting spectrum. Do such conditions and the resulting characteristic spectra really exist in astrophysical situations ?

\subsection{Toward a realistic description of the relativistic shock acceleration}  
  
Studies of - as far as possible - realistic conditions for the relativistic shock acceleration were  presented in a series of papers by Niemiec et al. (Niemiec \& Ostrowski 2004, 2006, Niemiec et al. 2006; see also Lemoine \ Pelletier 2006). We assumed 3-D static magnetic field perturbations upstream of the shock by imposing a large number of sinusoidal waves with different power spectra, $F(k)$. The flat spectrum, $F(k) \propto k^{-1}$, or the Kolmogorov spectrum, $F(k) \propto k^{-5/3}$, were considered in the wide wave-vector range ($k_{min}$, $k_{max}$), with $k_{max}/k_{min} = 10^5$. The downstream magnetic field structures were computed by respective compression of the upstream field at the shock. In the last of above papers an additional component of large amplitude short-wave MHD turbulence was assumed to be produce at the shock. The accelerated particle spectra were derived using the Monte Carlo simulations for a wide range of shock Lorentz factors - between 2 and 30 - and a selection of the mean magnetic field configurations and perturbation amplitudes.

\begin{figure}[hbt]  
\vspace{65mm}
\includegraphics{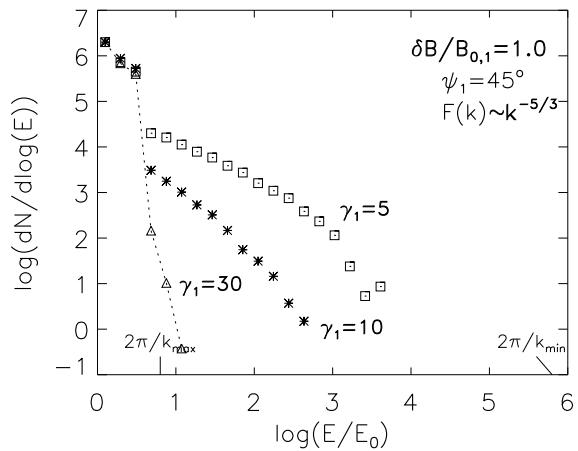}   
\vspace{0mm}
\caption{ 
The particle spectra derived  for superluminal relativistic shocks  with Lorentz factors $\gamma_1$ = 5, 10 and 30 (Niemiec \& Ostrowski 2006a).} 
\end{figure} 

\begin{figure*}[hbt]  
\vspace{75mm}
\includegraphics{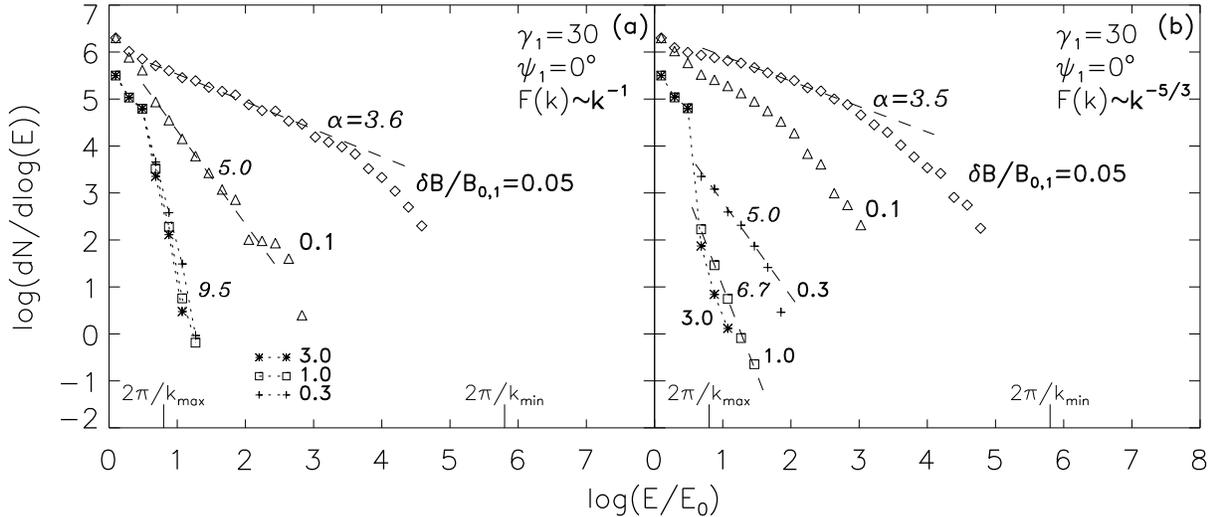}   
\vspace{0mm}\caption{ 
Particle spectra derived for parallel relativistic shocks  with Lorentz factor $\gamma_1$ =  30 (Niemiec \& Ostrowski 2006). For description see Fig.~5 and the original paper.} 
\end{figure*}

Let us take a look at a few characteristic results of these modelings. At Fig.~5 results derived for subluminal oblique mildly relativistic shocks are presented. One may note that introducing the energy scales to the modelling, in our units $2\pi / k_{max}$ and $2\pi / k_{min}$, lead to deviations of the resulting spectra from the power-law shape. The spectra, as expected from the previous discussion, are very flat for small amplitude turbulence, but steepen at larger amplitudes. An interesting feature is seen, that at very high particle energies, above the resonance range ($E > 2\pi/k_{min})$ the `short' waves weaker influence particle trajectories leading to the hard energy tails before the cut-offs imposed by the modelling.

When we considered oblique superluminal shocks the spectra consisted of the shock compressed injected component and the limited high energy tail until a cut-off well within the `resonance range' ($2\pi/k_{max}$, $2\pi/k_{min}$). As illustrated at Fig.~6 the tails fast diminish with the growing shock Lorentz factor, leaving for large $\gamma_1$ the `compressed component' only.

There are a few general observations for the first-order Fermi acceleration processes from these series of models. For particles in a low energy range of the resonance energies for the considered field perturbations the acceleration processes proceed in an ensemble of different oblique shocks, where each {\em local} mean magnetic field structure is formed as a superposition of the mean magnetic field $B_{0,1}$ and long wave field perturbations. Thus, there can occur significant differences between spectra generated in the presence of flat and steep turbulence power spectra, and the spectral indices significantly vary with the perturbations amplitudes. In parallel shocks the long wave perturbations introduce the acceleration effects observed with oblique magnetic fields (cf. also Ostrowski 1993). Thus in the ultrarelativistic parallel shocks propagating in highly turbulent medium these effects can lead to formation of particle distributions with cut-offs at relatively low energies, like in shocks with perpendicular field configurations (Fig.~7). If the shock wave generates a large amplitude short-wave turbulence downstream of the shock, the acceleration process can form a more extended power-law tail, but at higher particle energies the mean magnetic field or the long wave magnetic field structures start to dominate in shaping particle trajectories and thus the acceleration process, leading to results like the ones described above. In any studied case we were not able to create the scale free conditions in the acceleration process, leading to the wide range power law distribution of accelerated particles. It was possible only in limited energy ranges and the forms of the spectra depended usually a lot on the considered background conditions.

\subsection{Observational constraints on the shock acceleration from Cyg A hotspots}

Constraints for the above theoretical derivations can be provided by precise observations of energetic particle emission from objects harbouring the relativistic shocks. Such study performed for hotspots of the Cygnus A radiosource (Stawarz et al. 2007) reveals significant deviations of the derived electron spectra from the `standard' shock spectra. The resulting spectral energy distribution of the hotspot D is presented at Fig.~8, showing both the extended synchrotron component and the inverse-compton (IC) one modelled for optical and X-ray measurements. Additionally, the low Spitzer IR points provide additional important constraint for the IC spectral component. In derivation of the relativistic electrons distribution these measurements allow for excluding a possibility of substantial absorption in the low frequency synchrotron spectrum and thus require very flat distribution of low energy electrons. Thus the intrinsic electron spectrum (Fig.~9) is composed of a very flat low energy sector, with the energy spectral index $s \approx 1.5$, followed above a break at $E \approx 1$~GeV with the steep ($s > 3$) high energy sector. 
A possible interpretation of the spectrum considered by us was to assume a mildly-relativistic shock acceleration in the jet dominated by the protons bulk kinetic energy. The protons are expected establish the characteristic break energy scale, $E_{br} \sim 1$~GeV, at the shock transition layer. The electrons (or pairs) below this energy are expected to be accelerated within this layer due to electron-proton collective interactions. Above $E_{br}$, either the first order Fermi process acting at the shock creates a steep spectrum, or the acceleration process proceeds downstream of the shock in the second-order Fermi process. The existing theoretical models do not allow to reject any of these alternatives.

\begin{figure}[hbt]  
\vspace{60mm}
\includegraphics{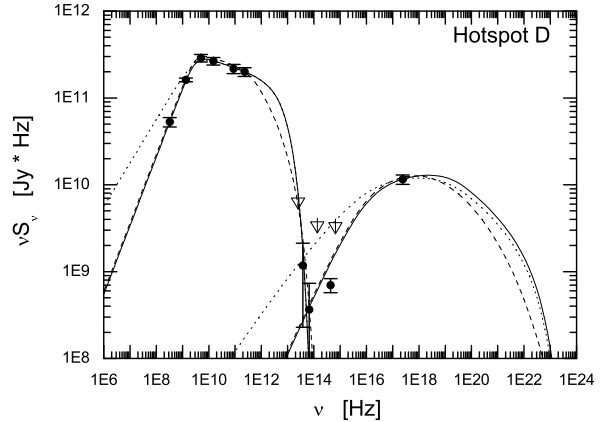}   
\vspace{0mm}
\caption{ 
Spectral energy distribution of emission from the hot spot D of Cyg A (Stawarz et al. 2007). One can clearly see both the synchrotron and the IC components. The Spitzer points in infrared and the optical point allow to exclude possibility of forming the measured flat low-frequency synchrotron component (dotted lines in synchrotron and IC spectral ranges) due to some self-absorption processes.} 
\end{figure} 
 
\begin{figure}[hbt]  
\vspace{60mm}
\includegraphics{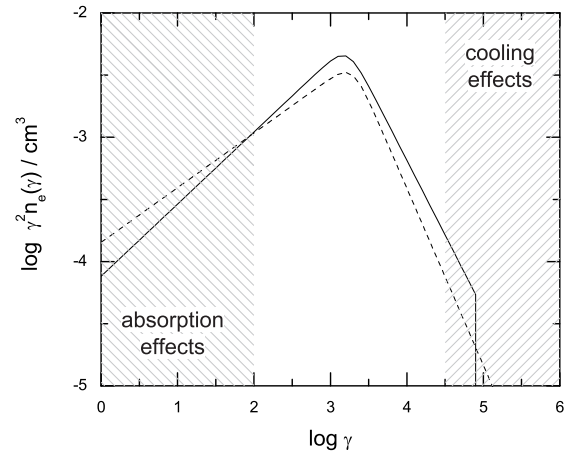}   
\vspace{0mm}
\caption{ 
Relativistic electron spectra in Cyg A hot-spots (Stawarz et al. 2007; here $\gamma$ is an electron Lorentz factor). Different spectral indices of the low energy and the high energy parts are expected to be intrinsic to the acceleration process, not the effect of the distribution `aging' downstream of the shock.} 
\end{figure}

\section{Energetic particle acceleration in shear layers and regions of relativistic turbulence}  
       
Acceleration processes acting, e.g., in AGN central sources and in shocks  
formed in large scale jets are not always able to explain the observed high  
energy electrons radiating away from the centre/shock. Among a few  
proposed possibilities explaining these data the relatively natural but  
still unexplored is the one involving particle acceleration within a shear layer transition at the interface between the jet and the surrounding medium. To date knowledge of physical conditions within such layers is very limited and only rough estimates for the considered acceleration processes are possible. Within the {\em subsonic} turbulent layer with a non-vanishing small velocity shear the 
ordinary second-order Fermi acceleration, as well as the process of `viscous' particle acceleration (cf. the review by Berezhko (1990) of the work done in early 80-th; Earl et al. 1988, Ostrowski 1990, 1998, 2000, Stawarz \& Ostrowski 2002) can take place. A mean particle energy gain per scattering in the later process scales as  
  
$$ {<\Delta E> \over E} \, \propto \left( { <\Delta U> \over c } \right)^2 \qquad ,   \eqno(4.1)$$   
  
\vspace{1mm}
\noindent  
where $< \Delta U >$ is the mean velocity difference between the  
`successive scattering acts'. It is proportional to the mean free path  
normal to the shear layer, $\lambda_n$, times the mean flow velocity  
gradient in this direction $ \nabla_n \cdot \vec{U} $. With $d$ denoting  
the shear layer thickness this gradient can be estimated as $| \nabla_n  
\cdot \vec{U} | \approx U/d$. Because the acceleration rate in the Fermi  
II process is $\propto (V / c)^2$ ($V \approx V_A$ is the wave velocity,  
$V_A$ -- the Alfv\'en velocity), the relative importance of both  
processes is given by a factor  
  
$$\left( {\lambda_n \over d} {U \over V} \right)^2 \qquad . \eqno(4.2)$$  
  
\vspace{1mm}
\noindent  
The relative efficiency of the viscous acceleration grows with $\lambda_n$ and in the formal limit of $\lambda_n \approx d$ and $V \ll c$ -- outside the equation (4.2) validity range -- it dominates over the Fermi acceleration to a large extent. Because accelerated particles can escape from the accelerating layer only due to a relatively inefficient radial diffusion, the formed particle spectra are expected be very flat up to the high energy cut-off, but the exact form of the spectrum depends on several unknown physical parameters of the boundary layer (Ostrowski 1998, 2000). 
     
For turbulent {\em relativistic} plasmas the second-order Fermi acceleration can in principle dominate over the viscous process at all particle energies. In the case of electrons the upper energy scale for the accelerated particles is provided by the radiation losses. A simple exercises with the above estimated acceleration scales and the synchrotron radiation loss scale yields - for the sources like small and large scale jets, radio hot spots or lobes -- the highest electron energies between 1 TeV and $10^3$ TeV, for the respective acceleration time scales $\sim 10^3$ s in ~gauss fields and up to $\sim 10^3$ yrs in $\mu$G fields (depending on the considered object). The much higher energy limits for protons are usually determined by the escape boundary conditions, not the radiative losses.
  
\section{Final remarks}  

A recent study of the first order Fermi acceleration processes at relativistic shocks, taking into account realistic assumptions about the physical conditions near the shock, reveals a few unexpected conclusions. The modelling of the acceleration process in mostly perpendicular (for relativistic velocities) shocks yields spectra consisting of the compressed injected part appended with a limited high energy tail. For given upstream conditions increasing the shock Lorentz  factor $\Gamma$ leads to steepening of such energetic tails, providing essentially the compressed injected component at large $\Gamma$. Another unexpected feature is observed dependence of the spectrum inclination on the turbulence amplitude also for the parallel shock waves and formation of cut-offs at such shocks for large $\Gamma$. Essentially no conditions studied by Niemiec et al. resulted in formation of the wide-range power-law particle distribution with {\em the universal spectral index 2.2}~. 
  
It can be of interest, with the presently published AUGER results (The Pierre Auger Collaboration 2007), that the modelling presented above seems to exclude the first-order Fermi acceleration at relativistic shocks as possible sources for highest energy particles registered in this experiment. 

In the same time the second-order Fermi processes acting in turbulent relativistic plasmas are expected to play significant role in cosmic ray acceleration. Thus, even facing substantial mathematical and physical difficulties, its deserve a detailed study. One may note that the acceleration processes accompanied the magnetic field reconnection processes are analogous to the second-order Fermi acceleration. Such processes always accompany the large amplitude MHD turbulence and generate turbulence. A few simple attempts to consider these processes were recently presented (e.g. Virtanen \& Vainio 2005).

When considering the relativistic shock acceleration one should also note  interesting approaches by Derishev et al. (2003) and Stern (2003), outside the classical Fermi scheme. They consider the acceleration processes in highly relativistic shocks or jet shear layers (Stern \& Poutanen 2006) involving particle-particle or particle-photon interactions at both sides of the shock.  
 
\subsection*{Acknowledgements}      
I am grateful to my collaborators Jacek Niemiec and {\L}ukasz Stawarz, whose significant work forms the main part of this rapport. The present work was supported by the Polish Ministry of Science and Higher Education in years 2005-2008 as a research project 1 P03D 003 29.

\end{document}